\documentclass[fullpaper,cameraready]{nldl}

\renewcommand{\DOI}{12345}

\usepackage[utf8]{inputenc}
\usepackage{url}
\usepackage{graphicx}
\usepackage{authblk}

\usepackage{algorithm}
\usepackage{algorithmic}

\usepackage{amsmath,diagbox}
\usepackage{colortbl}

\usepackage{multirow,makecell,amsmath,url, stfloats,enumitem}

\usepackage[square,numbers]{natbib}

\usepackage{xcolor}

\title{Multi-input segmentation of damaged brain in acute ischemic stroke patients using slow fusion with skip connection}
\author[1]{Luca Tomasetti\thanks{Corresponding Author: luca.tomasetti@uis.no}}
\author[1]{Mahdieh Khanmohammadi}
\author[1]{Kjersti Engan}
\author[1,2]{Liv Jorunn H\o{}llesli}
\author[1,2]{Kathinka D{\ae}hli Kurz}
\affil[1]{\small{Department of Electrical Engineering and Computer Science, University of Stavanger, 4021 Stavanger, Norway}}
\affil[2]{\small{Stavanger Medical Imaging Laboratory (SMIL), Department of Radiology, Stavanger University Hospital, 4019 Stavanger, Norway}}

\date{\vspace{-5ex}}

\begin{document}
\nldlmaketitle
\newcommand{\highlighti}[2]{%
\setlength{\fboxrule}{2pt}%
\fcolorbox{#1}{white}{#2}}

\begin{abstract}
Time is a fundamental factor during stroke treatments. A fast, automatic approach that segments the ischemic regions helps treatment decisions. In clinical use today, a set of color-coded parametric maps generated from computed tomography perfusion (CTP) images are investigated manually to decide a treatment plan.
We propose an automatic method based on a neural network using a set of parametric maps to segment the two ischemic regions (core and penumbra) in patients affected by acute ischemic stroke.
Our model is based on a convolution-deconvolution bottleneck structure with multi-input and slow fusion.
A loss function based on the focal Tversky index addresses the data imbalance issue.
The proposed architecture demonstrates effective performance and results comparable to the ground truth annotated by neuroradiologists.
A Dice coefficient of 0.81 for penumbra and 0.52 for core over the large vessel occlusion test set is achieved.
The full implementation is available at: \url{https://git.io/JtFGb}.
\end{abstract}

\section{Introduction}

A cerebral stroke is the second most common cause of death among adults worldwide \cite{wang2016global}.
Cerebral stroke can be divided into two general categories: ischemic and hemorrhagic stroke.
Ischemic stroke approximately represents 80\% of the totality of the strokes \cite{ojaghihaghighi2017comparison}.
The ischemic brain tissue is divided into two distinct regions during an ischemic stroke: the ischemic core (infarcted tissue) and the ischemic penumbra, a hypoperfused but viable tissue region.
Fast and correct visualization of the salvageable penumbra and the irreversibly damaged core tissues can benefit medical doctors for treatment planning in acute stroke patients (AIS).

Computed Tomography (CT) or Magnetic Resonance Imaging (MRI) are the two of the modalities used to diagnose acute stroke patients \cite{european2008guidelines}.
CT is preferred in many centers due to its high sensitivity for detecting hemorrhage, rapid scan times, and widespread availability.
Information about clinical severity are calculated, at hospital admission, using the National Institutes of Health Stroke Scale (NIHSS), and color-coded parametric maps (2D PMs) are generated using the 4D CT Perfusion (CTP) imaging usually performed immediately after hospital admission.
PMs are estimated to evaluate the changes in the tissue density over the injection of a contrast agent over time.
Time-to-peak (TTP), time-to-maximum ($\text{T}_{\text{Max}}$), cerebral blood flow (CBF), and cerebral blood volume (CBV), are all examples of PMs, derived from pixel information of a time density curve, generated from a CTP study \cite{kurz2016radiological}. Also, the maximum intensity projection (MIP) is found as the maximum Hounsfield unit value over the time sequence of the CTP providing a 3D volume.
In addition to diagnosing acute stroke, CT is also necessary for treatment decisions, with CTP being an essential modality with the ability to assess the penumbra and core.

Deep neural network (DNN) models have been proven to be an effective and beneficial tool for classification and segmentation tasks in many medical image analysis applications.
Various research groups have focused their effort on the study of ischemic strokes, and some methods have been developed for classifying and segmenting the infarct core \cite{abulnaga2018ischemic,clerigues2019acute,kasasbeh2019artificial,lucas2018multi}.
These methods rely on a set of PMs as input and ground truth images generated through follow-up images acquired hours after the stroke onset.
Nevertheless, the methods mentioned above only segment \emph{core} regions.
However, it could be more beneficial to acquire knowledge also about the penumbra regions since it is crucial for the treatment decisions during the first stages of the ischemic stroke \cite{murphy2006identification,tomasetti2021}.
To the best of our knowledge, Tomasetti et al. was the first research group to segment both \emph{core and penumbra} using machine learning and deep learning approaches \cite{tomasetti2020cnn,tomasetti2021}.

It is highly time-consuming to collect and label medical data, and transfer learning is a popular approach to solve problems related to medical images \cite{alzubaidi2021novel,wetteland2020multiscale}.
Over the past years, there have been numerous examples of transfer learning architectures used for various tasks in disparate domains pre-trained with the ImageNet dataset \cite{carreira2017quo,wetteland2020multiscale}.
Additionally, early and slow fusion approaches, with or without inflation, have been proven to improve accuracy in video classification \cite{carreira2017quo,karpathy2014large}, and medical diagnosis \cite{lalonde2019inn,meinich2020activity}.
An early fusion approach combines input information at the beginning of the process, allowing a network to increase the performances of the system using cross-correlation between data \cite{gadzicki2020early}.
The slow fusion approach slowly merges input information throughout the model permitting higher layers to access more global information \cite{karpathy2014large}.
An inflating technique has been proposed to use pre-trained weights from image classification networks in video classification models, expanding the filters from 2D (image-based) to 3D (video-based) \cite{carreira2017quo}.

\begin{figure*}[h!]
\centering
\includegraphics[width=\linewidth]{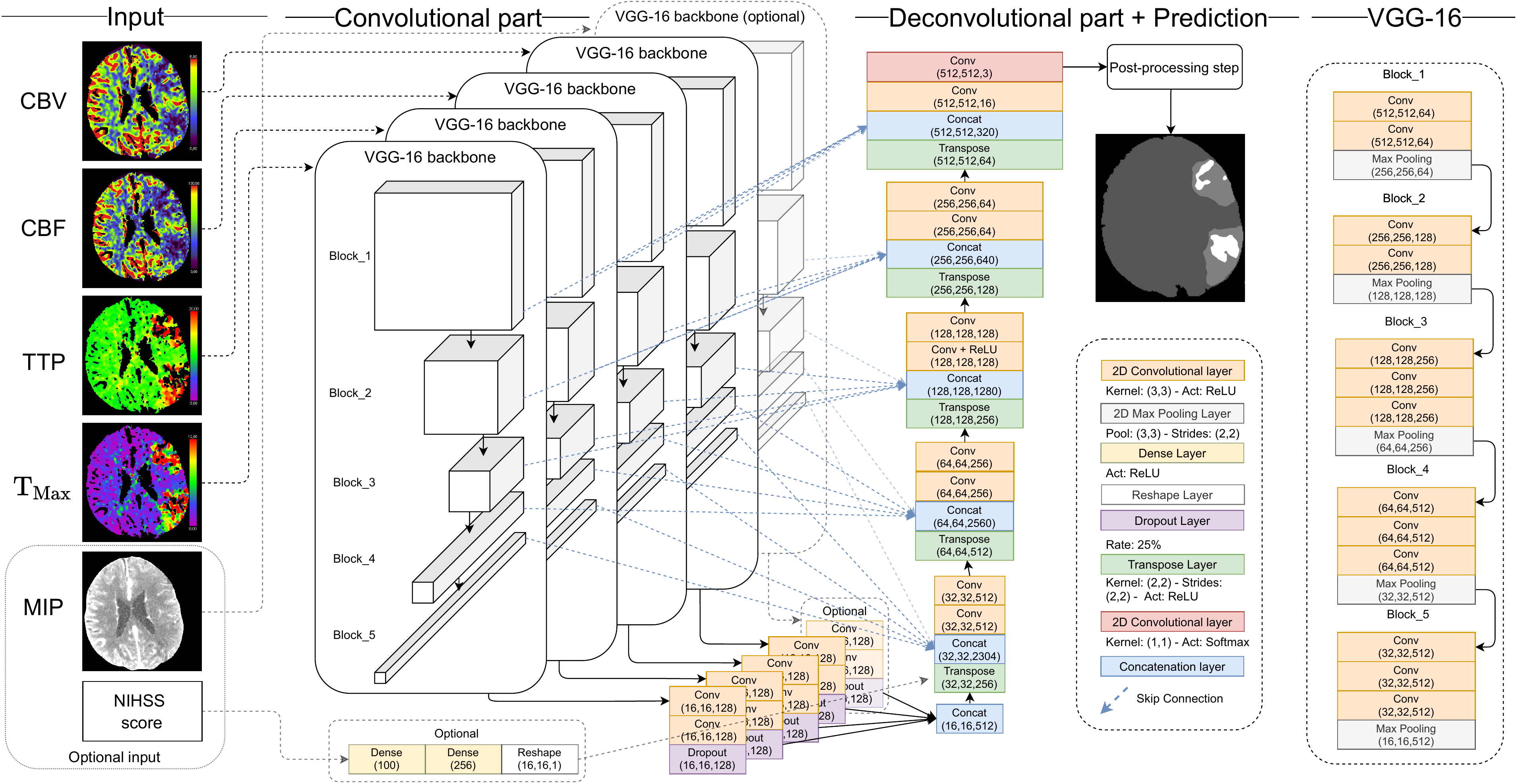}
\caption{Proposed architecture with multi-input and slow fusion.
The network has a convolutional part composed of four to five VGG-16 models.
Skip connections combined with feature concatenation from VGG-16 models ensure slow data fusion.
On the left, an overview of the feature extraction part of the VGG-16 architecture.
}
\label{fig:nn}
\end{figure*}

A proper understanding of \emph{both} ischemic regions is a major requisite for initial treatment decisions; however, it has not been fully explored in the previous researches; thus, in this work, we propose a DNN architecture to simultaneously segment both core and penumbra regions in AIS patients. We implemented a structure that was inspired by a multi-scale model proposed by Wetteland et al. \cite{wetteland2020multiscale}; however, while Wetteland's model used the same image but with different magnifications as input, we propose a different approach.
Our model uses a multi-input CNN with slow fusion, based on transfer learning from VGG-16 models pre-trained on ImageNet.
We want to investigate if the usage of the PMs in combination with MIP volume and NIHSS as input can produce meaningful results in the segmentation of ischemic regions, as this is already calculated and in use in clinical settings.
The paper contributes with:
\begin{itemize}[noitemsep,topsep=0pt,left=0pt]
\item A fully-automatic DNN method (Fig. \ref{fig:nn}) to segment both \emph{core and penumbra}, using color-coded PMs
acquired shortly after hospital admission when an AIS is expected.
\item A slow fusion multi-input approach is tested combining the PMs with other images and/or patient information.
\item The model is trained and tested using a dataset of patients affected by different levels of vessel occlusion for generalizing the input.
\item Manual annotations based on experts' assessment of the CTP with PMs and MIP are used as ground truth.
\end{itemize}

\section{Dataset}\label{sec:dat}

The study was approved by the Regional ethic committee project 2012/1499.
The dataset included CTP scans from 152 patients collected at Stavanger University Hospital (SUS) between January 2014 and August 2020. NIHSS score was available for all patients.
Based on the level of vessel occlusion using CT angiography, large vessel occlusion (LVO) was defined as occlusion of a large, proximal artery. 
Non-LVO was defined as patients with perfusion deficits with occlusion of a smaller, more distal artery or with perfusion deficits without visible artery occlusion.
Patients were divided into three groups based on the vessel occlusion severity: 77 patients with LVO, 60 Non-LVO patients, and 15 without ischemic stroke (WIS).

Two neuroradiologists with 16 and 3.5 years of clinical experience delineated ground truth core and penumbra regions in an in-house developed software tool. These delineations were based on visual information in the CBF, CBV, TTP, $\text{T}_{\text{Max}}$ and MIP images acquired directly after the CTP at hospital admission.
PMs and MIP have a $512\times 512$ pixel resolution, displaying only the brain region. Each patient study contains a number of brain slices between 13 and 27 for each scan.
Furthermore, the MRI examination performed within one to three days after the CT examination was studied and used in assistance to generate the ground truth images since follow-up MRI exhibits the final infarct core.
The dataset was randomly split into training (58\%), validation (20\%), and testing sets (22\%). The training set resulted in 89 patients (42 LVO, 38 Non-LVO, 9 WIS); the validation set resulted in 30 patients (16 LVO, 11 Non-LVO, 3 WIS), and the remaining 33 patients for testing (19 LVO, 11 Non-LVO, 3 WIS).

\section{Method}

The baseline model takes as input four PMs (CBF, CBV, TTP, $\text{T}_{\text{Max}}$) of each 2D brain slice, MIP images and, for some experiments, we also use the patient's NIHSS score as input.
The slow fusion is done both by concatenating the feature vectors from the different inputs and introducing skip connections from the different levels of the convolutional part.

The architecture presents a structure that resembles the U-Net model \cite{ronneberger2015u}.
Fig. \ref{fig:nn} displays a representation of the architecture.
The convolutional part of the network is based on four to five distinct VGG-16 networks.
This part is used to extract low-resolution features.
A detailed overview of a single VGG-16 architecture is displayed on the right of Fig. \ref{fig:nn}.
At the end of each VGG-16 network, two convolutional layers followed by a dropout layer are inserted to regularize the model.
The VGG-16 models were pre-trained with the ImageNet dataset.
The deconvolutional section is a series of transpose layers followed by convolutional layers.
The transpose layers receive in input a concatenation of the previous layer and the skip connection of the last convolutional layer for each block in the VGG-16 architectures.
This step is performed through a concatenation layer, where various inputs are concatenated together along the channel axis, providing a slow fusion of the low-level features and the features of the VGG-16 models.
All the convolutional layers use a ReLU activation function, except the last layer, which uses a softmax activation function that generates a probability vector for the three classes involved (i.e., core, penumbra, and healthy brain).

The model's output is a single 2D brain slice image with the same height and width dimensions as the input PMs.
The 3D volume is generated by concatenating all the 2D brain slice images sequentially.
To evaluate the accuracy of the predicted outputs, we use three distinct metrics: the Dice Coefficient, the Hausdorff distance, and the difference in volume among the predictions ($V_p$) and the ground truth images ($V_g$): $\Delta V = |V_g- V_p|$.

\subsection{Class imbalance \& Loss function}

The dataset has imbalanced classes: 93.1\% of the pixels belong to the healthy brain class, 6.2\% penumbra, and the remaining 0.7\% core.
This issue is even more pronounced in the Non-LVO group, where 0.2\% is core, 2.2\% penumbra, and the remaining 97.6\% belongs to healthy brain tissue.
To overcome this issue, we build our model to focus on two aspects: the \emph{loss function} and the \emph{Non-LVO group}.

A generalized focal loss, based on the Tversky index (TI) \cite{abraham2019novel,salehi2017tversky} was adopted.
The TI is a generalization of the Dice similarity coefficient.
The selected loss function was developed to address the data imbalance problem in medical image segmentation, improving the trade-off between precision and recall when training on small structures.
$\gamma, \alpha$, and $\beta$ are hyper-parameters of the Focal Tversky loss (FTL) \cite{abraham2019novel}.
Furthermore, during training, we emphasized the misclassification of penumbra and core class, in patients in the Non-LVO group, with a higher penalty in the loss because of the evident imbalance among classes in this sub-group.
A post-processing step is performed before generating the predicted outcome: a binary mask of the entire brain slice is created based on the MIP image to force the segmentation inside a valid area (the brain tissue).
Subsequently, from the softmax activation function, the highest probability value for each pixel was selected.

\section{Experiments \& Results}
\begin{figure*}[h!]
\centering
\includegraphics[width=\linewidth]{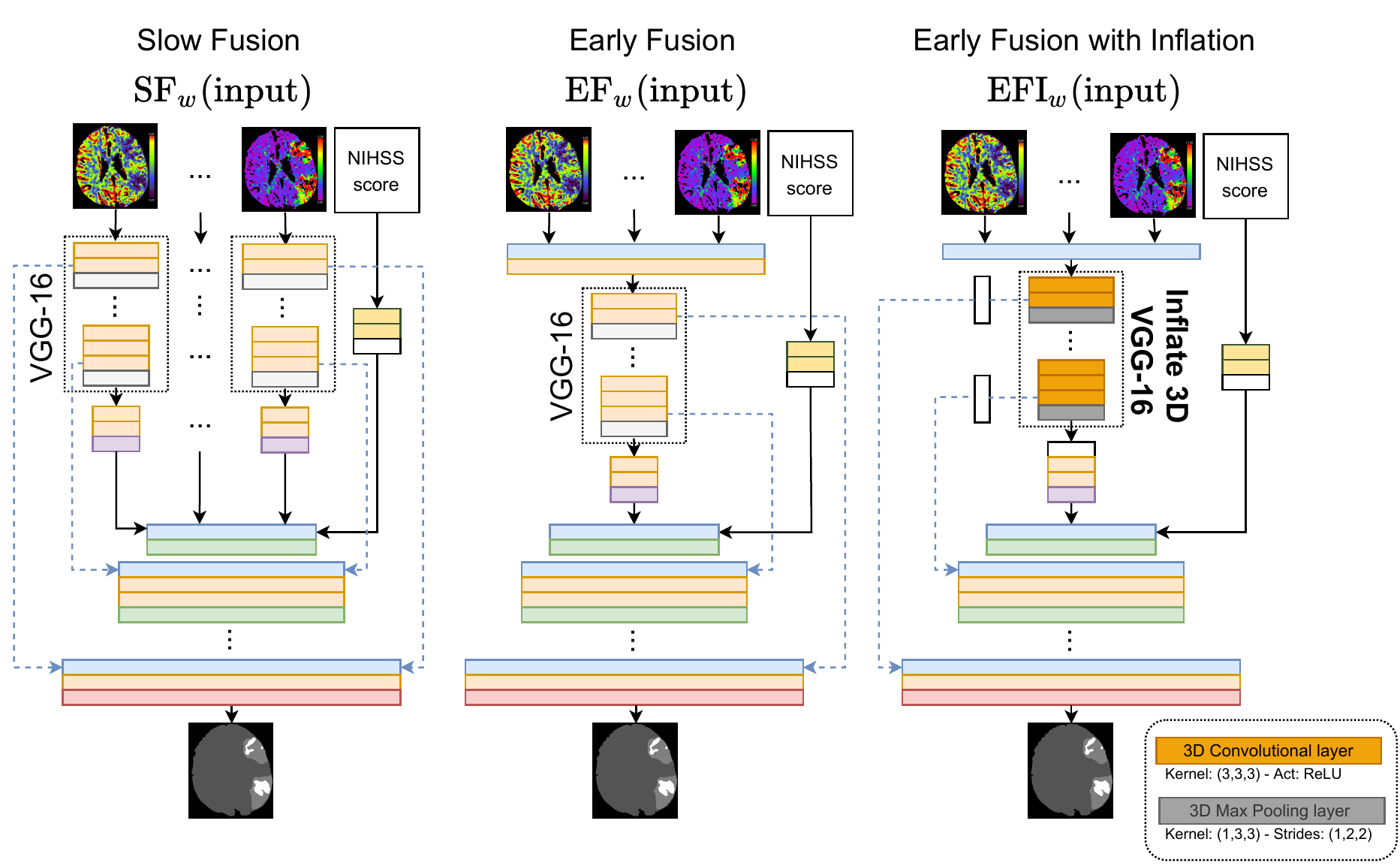}
\caption{Overview of the three models used for comparison in \textit{Exp-3}: the early fusion model ($\text{EF}_w$(\emph{input}));
the early fusion with inflation ($\text{EFI}_w$(\emph{input}));
$\text{SF}_w$(\emph{input}) is described in Fig. \ref{fig:nn}.}
\label{fig:coninf}
\end{figure*}

In the reminder of the paper we define the models as: $\text{SF}_w$(\emph{input}), where $w \in$\{\textbf{F}rozen, \textbf{U}nfrozen, \textbf{G}radual fine-tuning\} and the input are combination of PMs, MIP images (M), and NIHSS score (N).
We use the FTL as the loss function for our network.
We perform three different experiments: \textit{Exp-1)} Hyper-parameter search for the poposed method (Fig. \ref{fig:nn}). \textit{Exp-2)} combination of inputs and freeze/unfreeze variations of VGG-16; \textit{Exp-3)} comparison of different input-fusion methods (Fig. \ref{fig:coninf}).
For all experiments, the same setting was used: Adam \cite{kingma2014adam} was used as the optimizer function.
The batch size was set to 2, and each model was trained for 1000 epochs.
The validation FTL was monitored, and an early stopping was invoked if there was no improvement after 25 consecutive epochs.

In \textit{Exp-1} a hyper-parameter search is done running a large number of hyper-parameter combinations (over the same model) for finding the optimal values for the given task.
We ran experiments with distinct values for FTL hyper-parameters $\gamma \in [1,3], \alpha \in [0,1]$, and $\beta \in [0,1]$ as shown in Table \ref{tab:hp}, where each value represents the average of the patients in the validation set for the different severity levels.

\begin{table}[h]
\caption{\textit{Exp-1}: hyper-parameters search for the FTL loss, model $\text{SF}_F$(PMs). Each value represents the average of the patients in the validation set for the different severity levels and their standard deviation (SD).
}
\centering
\resizebox{.9\linewidth}{!}{
\begin{tabular}{c|c|c|c|cc|cc}
\hline
\multirow{3}{*}{Model} & \multicolumn{3}{c|}{\multirow{2}{*}{Parameters}} & \multicolumn{4}{c}{Dice Coefficient (Avg.) $\pm$ SD} \\
\cline{5-8}
& \multicolumn{3}{c|}{} & \multicolumn{2}{c|}{LVO} & \multicolumn{2}{c}{Non-LVO} \\
\cline{2-8}
& $\gamma$ & $\alpha$ & $\beta$ & Penumbra & Core & Penumbra & Core \\
\hline
\multirow{16}{*}{\rotatebox[origin=c]{90}{$\text{SF}_F$(PMs)}} & \multirow{4}{*}{1} & 1 & 1 & 0.68$\pm$0.2 & 0.29$\pm$0.3 & \textbf{0.30$\pm$0.3} & 0.13$\pm$0.2 \\ \cline{3-4}
& & 0.3 & 0.7 & 0.36$\pm$0.2 & 0.37$\pm$0.3 & 0.06$\pm$0.1 & 0.21$\pm$0.3 \\ \cline{3-4}
& & 0.5 & 0.5 & 0.66$\pm$0.2 & 0.35$\pm$0.3 & \textbf{0.30$\pm$0.4}& 0.19$\pm$0.2 \\ \cline{3-4}
& & 0.7 & 0.3 & 0.68$\pm$0.2 &0.31$\pm$0.3& 0.28$\pm$0.3& 0.17$\pm$0.2 \\ \cline{2-8}
& \multirow{3}{*}{4/3} & 0.3 & 0.7 & 0.68$\pm$0.2 &0.35$\pm$0.3 &0.29$\pm$0.4&0.17$\pm$0.2 \\ \cline{3-4}
& & 0.5 & 0.5 & 0.69$\pm$0.2 &0.36$\pm$0.3 & \textbf{0.30$\pm$0.3} &0.17$\pm$0.3 \\ \cline{3-4}
& & 0.7 & 0.3 & \textbf{0.71$\pm$0.1}&0.37$\pm$0.3&0.27$\pm$0.3&\textbf{0.22$\pm$0.3}\\ \cline{2-8}
& \multirow{3}{*}{1.5} & 0.3 & 0.7 & 0.67$\pm$0.2&0.30$\pm$0.3&0.27$\pm$0.3&0.14$\pm$0.2\\ \cline{3-4}
& & 0.5 & 0.5 & 0.67$\pm$0.2& 0.30$\pm$0.3 & 0.29$\pm$0.3 & 0.14$\pm$0.2 \\ \cline{3-4}
& & 0.7 & 0.3 & 0.70$\pm$0.3&0.37$\pm$0.2 &0.29$\pm$0.3&0.20$\pm$0.3 \\ \cline{2-8}
& \multirow{3}{*}{2} & 0.3 & 0.7 & 0.67$\pm$0.2& 0.00$\pm$0.0&\textbf{0.30$\pm$0.4}&0.00$\pm$0.0\\ \cline{3-4}
& & 0.5 & 0.5 & 0.70$\pm$0.2&0.00$\pm$0.0 &0.28$\pm$0.3&0.00$\pm$0.0\\ \cline{3-4}
& & 0.7 & 0.3 & 0.70$\pm$0.2&\textbf{0.39$\pm$0.3}&0.29$\pm$0.3&0.20$\pm$0.3\\ \cline{2-8}
& \multirow{3}{*}{3} & 0.3 & 0.7 & 0.00$\pm$0.0 &0.00$\pm$0.0&0.00$\pm$0.0&0.00$\pm$0.0 \\ \cline{3-4}
& & 0.5 & 0.5 & 0.70$\pm$0.2&0.00$\pm$0.0 & \textbf{0.30$\pm$0.4} & 0.00$\pm$0.0\\ \cline{3-4}
& & 0.7 & 0.3 & 0.68$\pm$0.2 & 0.37$\pm$0.3&0.29$\pm$0.3&0.19$\pm$0.3 \\
\hline
\end{tabular}}
\label{tab:hp}
\end{table}

\begin{table*}[b]
\centering
\caption{Statistical results over the validation set for the models divided for \textit{Exp-2} and \textit{Exp-3}.
The last two rows contain results for \textit{Exp-3}.
Dice coefficient, Hausdorff distance, and the volume difference are the metrics considered to select the best model. Each value exhibits the average of the patients in the validation set and the standard deviation (SD) for the distinct groups.
For the metrics: $\Uparrow$ indicates that higher values are better, while with $\Downarrow$ lower values are preferable.
Highlighted values represent the best result for a specific class and metric.
The selected model is highlighted inside a red rectangle.
}
\label{tab:exps}%
\resizebox{\linewidth}{!}{
\begin{tabular}{c|ccc|c|c|c|c|c|c|c}
\hline
\multirow{3}{*}{Model} & \multicolumn{3}{c|}{\multirow{2}{*}{Input}} & \multicolumn{2}{c|}{Dice Coeff. (Avg.) $\pm$ SD $\Uparrow$} & \multicolumn{2}{c|}{Hausdorff Dist. (Avg.) $\pm$ SD $ \Downarrow$} & \multicolumn{3}{c}{$\Delta V$ (Avg.) $\pm$ SD (ml) $\Downarrow$} \\
\cline{5-11}
& & & & LVO & Non-LVO &LVO & Non-LVO & LVO & Non-LVO & WIS \\ \cline{2-11}
& PMs & MIP & NIHSS & \multicolumn{3}{c}{Penumbra} & \multicolumn{1}{c}{\diagbox[height=1em]{}{}} & \multicolumn{3}{c}{Core} \\
\hline\hline
\multicolumn{11}{c}{\textit{Exp-2} - Layer weights: \textbf{Frozen}} \\
\hline
$\text{SF}_F$(PMs) & X & & & \diagbox[height=2.19em]{0.71$\pm$0.1}{\textbf{0.37$\pm$0.3}}&\diagbox[height=2.19em]{0.27$\pm$0.3}{0.22$\pm$0.3}& \diagbox[height=2.19em]{5.9$\pm$1.0}{2.7$\pm$1.8}&\diagbox[height=2.19em]{3.2$\pm$1.5}{0.9$\pm$0.8}&\diagbox[height=2.19em]{27.0$\pm$28}{9.4$\pm$20}&\diagbox[height=2.19em]{10.0$\pm$15}{0.8$\pm$1.3}&\diagbox[height=2.19em]{9.8$\pm$8.0}{0.5$\pm$0.5} \\
\cline{1-1} \cline{5-11}
$\text{SF}_F$(PMs,M) & X & X & & \diagbox[height=2.19em]{0.69$\pm$0.2}{0.36$\pm$0.3}& \diagbox[height=2.19em]{0.29$\pm$0.3}{0.20$\pm$0.3} & \diagbox[height=2.19em]{5.9$\pm$0.9}{3.0$\pm$1.7} & \diagbox[height=2.19em]{3.1$\pm$1.4}{1.0$\pm$0.8} & \diagbox[height=2.19em]{25.9$\pm$27}{10.0$\pm$21} &\diagbox[height=2.19em]{5.5$\pm$6.0}{0.8$\pm$1.0} & \diagbox[height=2.19em]{8.1$\pm$7.0}{0.5$\pm$0.3} \\
\cline{1-1} \cline{5-11}
$\text{SF}_F$(PMs,N) & X & & X & \diagbox[height=2.19em]{0.70$\pm$0.2}{0.36$\pm$0.3} & \diagbox[height=2.19em]{0.29$\pm$0.3}{0.16$\pm$0.2} & \diagbox[height=2.19em]{5.6$\pm$1.2}{\textbf{2.3$\pm$1.9}} & \diagbox[height=2.19em]{2.3$\pm$1.3}{0.7$\pm$0.6} & \diagbox[height=2.19em]{27.0$\pm$37}{5.0$\pm$9.0} &\diagbox[height=2.19em]{4.1$\pm$4.8}{0.6$\pm$1.2} &\diagbox[height=2.19em]{4.0$\pm$3.4}{0.1$\pm$0.1} \\
\cline{1-1} \cline{5-11}
$\text{SF}_F$(PMs,M,N) & X & X & X & \diagbox[height=2.19em]{0.68$\pm$0.2}{0.34$\pm$0.3} & \diagbox[height=2.19em]{0.3$\pm$0.3}{0.18$\pm$0.3} &\diagbox[height=2.19em]{5.7$\pm$1.4}{2.5$\pm$1.8}& \diagbox[height=2.19em]{2.4$\pm$1.2}{0.8$\pm$0.7} & \diagbox[height=2.19em]{29.9$\pm$37}{6.3$\pm$12} & \diagbox[height=2.19em]{\textbf{2.8$\pm$3.0}}{0.5$\pm$1.0} & \diagbox[height=2.19em]{3.9$\pm$4.0}{\textbf{0.0$\pm$0.0}} \\
\hline\hline
\multicolumn{11}{c}{\textit{Exp-2} - Layer weights: \textbf{Unfrozen}} \\
\hline
$\text{SF}_U$(PMs) & X & & & \diagbox[height=2.19em]{0.70$\pm$0.2}{0.34$\pm$0.3} &\diagbox[height=2.19em]{0.29$\pm$0.4}{0.24$\pm$0.3} & \diagbox[height=2.19em]{5.4$\pm$1.3}{2.7$\pm$1.7} & \diagbox[height=2.19em]{2.1$\pm$1.5}{0.7$\pm$0.7} & \diagbox[height=2.19em]{29.8$\pm$36}{6.5$\pm$14} & \diagbox[height=2.19em]{4.2$\pm$6.9}{0.5$\pm$0.7}&\diagbox[height=2.19em]{\textbf{0.1$\pm$0.1}}{\textbf{0.0$\pm$0.0}} \\
\cline{1-1} \cline{5-11}
$\text{SF}_U$(PMs,M) & X & X & & \diagbox[height=2.19em]{0.70$\pm$0.2}{0.36$\pm$0.3} & \diagbox[height=2.19em]{0.34$\pm$0.3}{\textbf{0.24$\pm$0.3}} & \diagbox[height=2.19em]{5.6$\pm$1.4}{2.5$\pm$1.8} & \diagbox[height=2.19em]{2.3$\pm$1.6}{0.6$\pm$0.6} & \diagbox[height=2.19em]{28.1$\pm$40}{6.5$\pm$8.1} & \diagbox[height=2.19em]{4.9$\pm$9.8}{0.4$\pm$0.8} & \diagbox[height=2.19em]{0.9$\pm$1.5}{\textbf{0.0$\pm$0.0}}\\
\cline{1-1} \cline{5-11}
$\text{SF}_U$(PMs,N) & X & & X & \diagbox[height=2.19em]{\textbf{0.72$\pm$0.2}}{0.36$\pm$0.3} & \diagbox[height=2.19em]{0.29$\pm$0.3}{0.23$\pm$0.3} & \diagbox[height=2.19em]{5.6$\pm$1.1}{2.6$\pm$1.8} & \diagbox[height=2.19em]{2.7$\pm$1.4}{0.9$\pm$0.7} & \diagbox[height=2.19em]{28.8$\pm$28}{5.2$\pm$10} &\diagbox[height=2.19em]{7.8$\pm$12}{0.5$\pm$0.7}&\diagbox[height=2.19em]{3.4$\pm$2.7}{0.1$\pm$0.1} \\
\cline{1-1} \cline{5-11}
$\text{SF}_U$(PMs,M,N) & X & X & X & \diagbox[height=2.19em]{0.71$\pm$0.2}{0.36$\pm$0.3} & \diagbox[height=2.19em]{0.32$\pm$0.3}{0.22$\pm$0.3} & \diagbox[height=2.19em]{5.7$\pm$1.0}{2.8$\pm$1.7} & \diagbox[height=2.19em]{2.4$\pm$1.6}{1.0$\pm$0.8} & \diagbox[height=2.19em]{29.0$\pm$25}{7.0$\pm$14} & \diagbox[height=2.19em]{9.0$\pm$18}{0.8$\pm$0.9} & \diagbox[height=2.19em]{5.5$\pm$4.5}{0.1$\pm$0.1} \\
\hline\hline
\multicolumn{11}{c}{\textit{Exp-2} - Layer weights: \textbf{Gradual Fine-tuning}} \\
\hline
$\text{SF}_G$(PMs) & X & & & \diagbox[height=2.19em]{0.71$\pm$0.2}{0.35$\pm$0.3} & \diagbox[height=2.19em]{0.30$\pm$0.3}{0.19$\pm$0.3} & \diagbox[height=2.19em]{5.8$\pm$1.0}{2.4$\pm$1.8} & \diagbox[height=2.19em]{3.0$\pm$1.4}{0.8$\pm$0.7} & \diagbox[height=2.19em]{29.3$\pm$25}{4.7$\pm$7.6} & \diagbox[height=2.19em]{11.8$\pm$22}{0.6$\pm$0.8} & \diagbox[height=2.19em]{6.2$\pm$5.5}{\textbf{0.0$\pm$0.0}} \\
\cline{1-1} \cline{5-11}
$\text{SF}_G$(PMs,M) & X & X & & \diagbox[height=2.19em]{0.69$\pm$0.2}{0.35$\pm$0.3} & \diagbox[height=2.19em]{\textbf{0.34$\pm$0.6}}{0.22$\pm$0.4} & \diagbox[height=2.19em]{5.4$\pm$1.4}{2.4$\pm$1.8} & \diagbox[height=2.19em]{\textbf{1.9$\pm$1.2}}{\textbf{0.6$\pm$0.5}} & \diagbox[height=2.19em]{27.8$\pm$38}{5.2$\pm$9.4} & \diagbox[height=2.19em]{3.3$\pm$3.7}{\textbf{0.3$\pm$0.3}} & \diagbox[height=2.19em]{2.4$\pm$3.8}{\textbf{0.0$\pm$0.0}} \\
\cline{1-1} \cline{5-11}
\highlighti{red}{$\text{SF}_G$(PMs,N)} & X & & X & \diagbox[height=2.19em]{\textbf{0.72$\pm$0.2}}{\textbf{0.37$\pm$0.3}} & \diagbox[height=2.19em]{0.31$\pm$0.3}{0.21$\pm$0.3} & \diagbox[height=2.19em]{\textbf{5.3$\pm$1.4}}{2.4$\pm$1.7} &\diagbox[height=2.19em]{2.3$\pm$1.4}{0.7$\pm$0.6}& \diagbox[height=2.19em]{26.0$\pm$35}{\textbf{4.4$\pm$7.0}} & \diagbox[height=2.19em]{6.5$\pm$14}{0.5$\pm$0.7} &\diagbox[height=2.19em]{1.0$\pm$1.7}{\textbf{0.0$\pm$0.0}} \\
\cline{1-1} \cline{5-11}
$\text{SF}_G$(PMs,M,N) & X & X & X & \diagbox[height=2.19em]{0.68$\pm$0.3}{0.34$\pm$0.3} & \diagbox[height=2.19em]{0.31$\pm$0.3}{0.18$\pm$0.3} & \diagbox[height=2.19em]{5.5$\pm$1.3}{2.5$\pm$1.8} & \diagbox[height=2.19em]{2.3$\pm$1.3}{0.7$\pm$0.6} & \diagbox[height=2.19em]{26.9$\pm$35}{5.1$\pm$9.6} & \diagbox[height=2.19em]{5.5$\pm$9.2}{0.5$\pm$0.8}& \diagbox[height=2.19em]{0.9$\pm$1.1}{\textbf{0.0$\pm$0.0}} \\
\hline \hline
\multicolumn{11}{c}{\textit{Exp-3} - Layer weights: \textbf{Gradual Fine-tuning}} \\
\hline
$\text{EF}_G$(PMs,N) & X & & X & \diagbox[height=2.19em]{0.68$\pm$0.2}{0.26$\pm$0.3} & \diagbox[height=2.19em]{0.32$\pm$0.3}{0.19$\pm$0.3} & \diagbox[height=2.19em]{5.6$\pm$1.4}{2.4$\pm$1.9} &\diagbox[height=2.19em]{2.3$\pm$1.5}{0.7$\pm$0.6} &\diagbox[height=2.19em]{31.8$\pm$47}{8.0$\pm$21} & \diagbox[height=2.19em]{5.3$\pm$10}{0.7$\pm$0.9} & \diagbox[height=2.19em]{1.8$\pm$0.3}{\textbf{0.0$\pm$0.0}} \\
\cline{1-1} \cline{5-11}
$\text{EFI}_G$(PMs,N) & X & & X & \diagbox[height=2.19em]{0.67$\pm$0.2}{0.29$\pm$0.3} & \diagbox[height=2.19em]{0.29$\pm$0.3}{0.16$\pm$0.2} & \diagbox[height=2.19em]{6.0$\pm$0.8}{2.5$\pm$1.9} & \diagbox[height=2.19em]{3.3$\pm$1.5}{0.7$\pm$0.6} & \diagbox[height=2.19em]{\textbf{25.5$\pm$28}}{5.3$\pm$10} & \diagbox[height=2.19em]{6.3$\pm$7.1}{0.6$\pm$1.2} &\diagbox[height=2.19em]{6.7$\pm$4.7}{\textbf{0.0$\pm$0.0}} \\
\hline
\end{tabular}}
\end{table*}

For \textit{Exp-2} we have combined our slow fusion multi-input baseline model with different combinations of inputs to understand if diversity in the input can improve training the model. The various combinations are shown in Table \ref{tab:exps} with the corresponding results.
From the baseline input (PMs), we added MIP images as input or the NIHSS score or combined both MIP images and NIHSS score.
Models for all these experiments were based on a multi-input with slow fusion.
VGG-16 models were trained with frozen weights, unfrozen weights, and a gradual fine-tuning approach for each experiment.
The latter setting was developed in three steps: first, the model was trained with all the VGG-16 weights frozen; secondly, after monitoring the validation loss having no improvements for 25 consecutive epochs, the bottom half of the weights were unfrozen, and the training continued; at last, when no improvement in the validation loss was detected again, all weights were unfrozen, the training continued, and the validation loss was monitored.
We have selected $\text{SF}_G$(PMs,N) as the proposed model.

\begin{table*}[h!]
\caption{Evaluation metrics for the selected method ($\text{SF}_G$(PMs,N)) over the 33 test patients (19 LVO, 11 Non-LVO, 3 WIS) manually annotated by two experts neuroradiologists ($\text{NR}_1$ and $\text{NR}_2$).
}\label{tab:iov}%
\centering
\resizebox{\linewidth}{!}{
\begin{tabular}{c|c|c|c|c|c|c|c|c|c|c}
\hline
\multirow{3}{*}{Model} &\multicolumn{3}{c|}{Dice Coefficient (Avg.) $\pm$ SD $\Uparrow$} & \multicolumn{3}{c|}{ Hausdorff Distance (Avg.) $\pm$ SD $\Downarrow$} & \multicolumn{4}{c}{$\Delta V$ (Avg.) $\pm$ SD (ml) $\Downarrow$} \\
\cline{2-11}
& LVO &Non-LVO & All & LVO & Non-LVO & All & LVO & Non-LVO &WIS &All \\ \cline{2-11}
& \multicolumn{4}{c}{Penumbra} & \multicolumn{2}{c}{\diagbox[height=1em]{}{}} & \multicolumn{4}{c}{Core} \\
\hline
\makecell{$\text{SF}_G$(PMs,N) \\ vs ($\text{NR}_1$ \& $\text{NR}_2$)} & \diagbox[height=2.19em]{0.81$\pm$0.1}{0.52$\pm$0.2} & \diagbox[height=2.19em]{0.48$\pm$0.3}{0.12$\pm$0.2} & \diagbox[height=2.19em]{0.63$\pm$0.3}{0.34$\pm$0.3} & \diagbox[height=2.19em]{5.0$\pm$1.2}{3.2$\pm$1.5} &\diagbox[height=2.19em]{2.2$\pm$0.9}{0.7$\pm$1.0} & \diagbox[height=2.19em]{3.7$\pm$2.0}{2.1$\pm$2.8} & \diagbox[height=2.19em]{15.5$\pm$12}{6.1$\pm$5.1} &\diagbox[height=2.19em]{4.4$\pm$4.4}{0.7$\pm$1.4} & \diagbox[height=2.19em]{0.2$\pm$0.4}{0.0$\pm$0.0} & \diagbox[height=2.19em]{10.5$\pm$11}{3.7$\pm$4.8} \\
\hline \hline
$\text{NR}_1$ vs. $\text{NR}_2$ & \diagbox[height=2.19em]{0.78$\pm$0.1}{0.44$\pm$0.2} &\diagbox[height=2.19em]{0.65$\pm$0.2}{0.15$\pm$0.2} & \diagbox[height=2.19em]{0.67$\pm$0.2}{0.30$\pm$0.3} & \diagbox[height=2.19em]{5.1$\pm$$\pm$1.0}{3.2$\pm$1.4} & \diagbox[height=2.19em]{1.9$\pm$1.4}{0.5$\pm$0.9} & \diagbox[height=2.19em]{3.6$\pm$2.2}{2.0$\pm$1.8} & \diagbox[height=2.19em]{33.3$\pm$28}{5.6$\pm$4.3} & \diagbox[height=2.19em]{5.5$\pm$9.2}{0.7$\pm$1.9} & \diagbox[height=2.19em]{0.0$\pm$0.0}{0.0$\pm$0.0} & \diagbox[height=2.19em]{21.0$\pm$26}{3.5$\pm$4.2} \\
\hline
$\text{SF}_G$(PMs,N) vs. $\text{NR}_1$ & \diagbox[height=2.19em]{0.84$\pm$0.1}{0.57$\pm$0.2} & \diagbox[height=2.19em]{0.48$\pm$0.3}{0.12$\pm$0.2} & \diagbox[height=2.19em]{0.64$\pm$0.3}{0.37$\pm$0.3} & \diagbox[height=2.19em]{4.9$\pm$1.1}{3.0$\pm$1.4} & \diagbox[height=2.19em]{2.2$\pm$1.0}{0.7$\pm$1.0} & \diagbox[height=2.19em]{3.6$\pm$1.9}{1.9$\pm$1.7} & \diagbox[height=2.19em]{11.2$\pm$11}{6.2$\pm$5.7} & \diagbox[height=2.19em]{4.4$\pm$4.4}{0.7$\pm$1.4} & \diagbox[height=2.19em]{0.3$\pm$0.4}{0.0$\pm$0.0}& \diagbox[height=2.19em]{7.9$\pm$9.4}{3.8$\pm$5.2}\\
\hline
$\text{SF}_G$(PMs,N) vs. $\text{NR}_2$ & \diagbox[height=2.19em]{0.78$\pm$0.1}{0.44$\pm$0.2} & \diagbox[height=2.19em]{0.43$\pm$0.3}{0.1$\pm$0.2} & \diagbox[height=2.19em]{0.59$\pm$0.3}{0.29$\pm$0.3} & \diagbox[height=2.19em]{5.4$\pm$1.1}{3.3$\pm$1.4} & \diagbox[height=2.19em]{2.7$\pm$1.4}{0.5$\pm$0.8} & \diagbox[height=2.19em]{4.0$\pm$2.1}{2.1$\pm$1.8} & \diagbox[height=2.19em]{29.2$\pm$30}{6.8$\pm$5.9} & \diagbox[height=2.19em]{8.2$\pm$9.6}{0.3$\pm$0.6} & \diagbox[height=2.19em]{0.3$\pm$0.4}{0.0$\pm$0.0} & \diagbox[height=2.19em]{19.6$\pm$26}{4.0$\pm$5.5} \\
\hline
\end{tabular}}
\end{table*}

To understand if using multi-input and slow fusion is suitable for this task, in \textit{Exp-3} we compared it with two models: an early fusion ($\text{EF}_G$(PMs,N)), and early fusion with inflation ($\text{EFI}_G$(PMs,N)), adopting the same multi-input idea but with different fusion approacheas.
The inflation approach converts 2D into 3D layers, adding a temporal dimension.
The inflation followed the idea of the I3D network by Carreira et al. \cite{carreira2017quo}, where they introduced video classification models with an inflated ImageNet pre-trained image classification architecture.
The setting and hyper-parameters of these two new architectures are the same as the selected model to maintain a fair comparison.
Fig. \ref{fig:coninf} presents an overview of the model architectures.
For $\text{EF}_G$(PMs,N), the four input parametric maps are concatenated together over the channel axis to generate a single input volume passing through a 2D convolutional layer to reduce the channel dimension to feed it to a single VGG-16 backbone.
Differently, $\text{EFI}_G$(PMs,N)'s inputs are concatenated over the time dimension; the filters and pooling kernels of the VGG-16 architecture are inflated, remodeling squared filters into cubic filters.
The bottom of Table \ref{tab:exps} shows the results of these two models in comparison with the other experiments performed.

$\text{SF}_G$(PMs,N) is seen as having good overall performance and is chosen as the proposed model.
The model is tested with a previously unseen test set, and the performance is compared with manual annotations from two different experts.
Table \ref{tab:iov} presents the test results and the inter-observer variability in comparison with two expert neuroradiologists.
Fig. \ref{fig:plots} shows examples of predictions from six patients from the test set using the proposed model.

\section{Discussion}

We developed a multi-input CNN with early and slow fusion using transfer learning in this work. The proposed network aims to simultaneously segment dead (core) and salvageable tissues (penumbra) in AIS patients.
The proposed method is learned on patients with or without ischemic stroke and for different vessel occlusion severities.
This generalization helps the model correctly segment most ischemic regions, regardless of the patient group.
After a series of experiments, multi-input including NIHSS, by a slow fusion approach, $\text{SF}_G$(PMs,N), is chosen as the proposed method based on the total of the results.


\begin{figure}[h!]
\centering
\includegraphics[width=\linewidth]{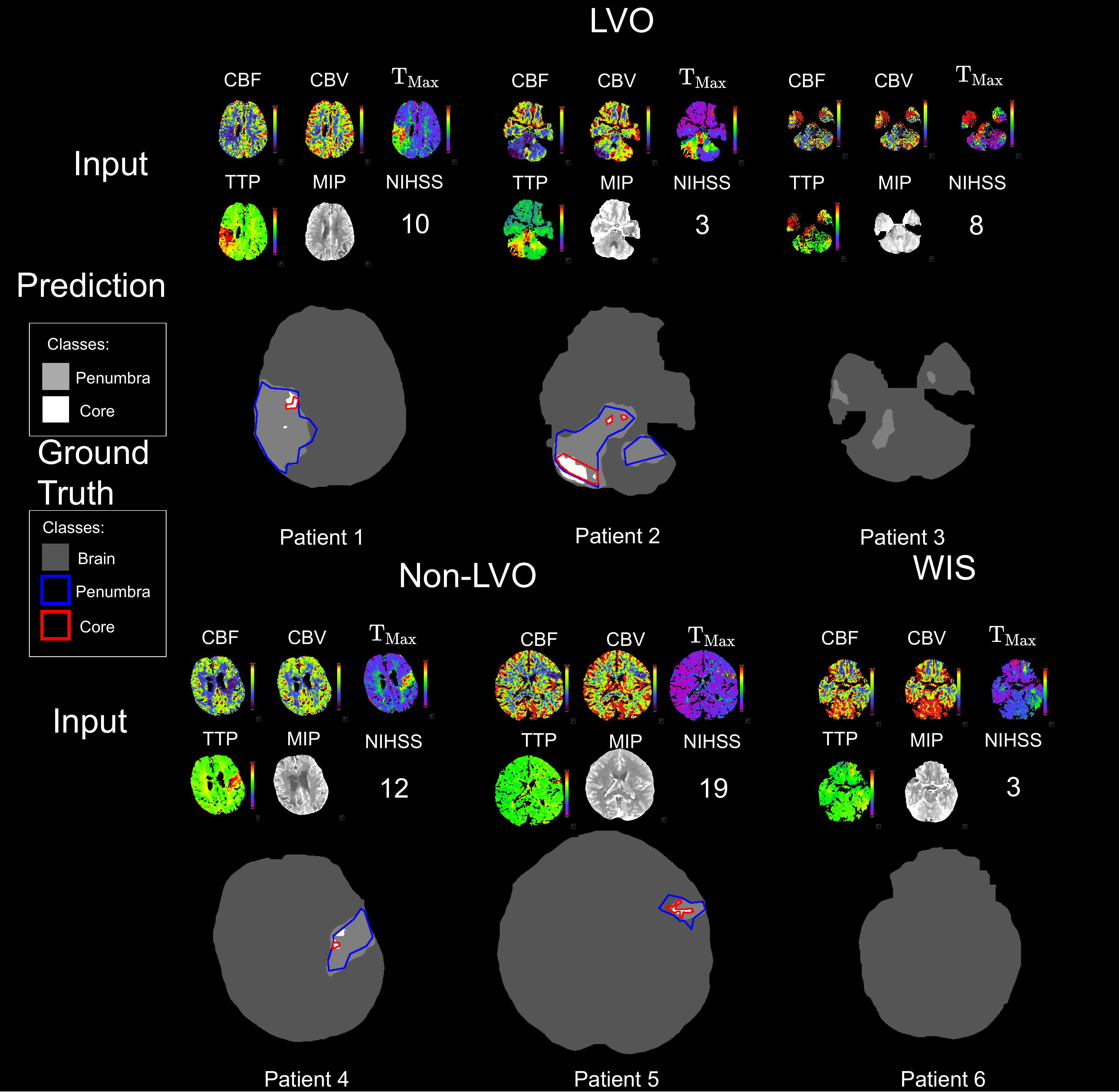}
\caption{Prediction results for six test patients using the selected model, the set of all possible inputs, and the relative ground truth images. }
\label{fig:plots}
\end{figure}

A hyper-parameter search was performed for the first experiment.
Table \ref{tab:hp} shows the average Dice coefficient for selecting the optimal hyper-parameters for the FTL function.
Our observations showed that $\gamma=4/3$, $\alpha=0.7$, and $\beta = 1-\alpha$ give satisfactory results for all the classes in the two groups.
Thus we apply these parameters in the following experiments.
From the validation results of Table \ref{tab:exps} it can be seen that unfreezing the weights of the VGG-16 feature extractors improves the models but gives an overestimation of the volume for both the classes and also that the gradual fine-tuning approach gives a slight improvement when compared to unfreezing all weights from the start.
The choice of freezing the parameters reduces training time since a smaller set of parameters needs to be learned; however, the statistical results are less than satisfactory; this could be since PMs are not included in the ImageNet dataset, then the weights are not optimized for these images.
Therefore, at the cost of a longer training time, fully unfreezing the weights or using a gradual fine-tuning technique will allow the model to familiarize and learn this particular dataset more accurately.

From the validation results of the different gradual fine-tuned models, it is clear that multi-input fusion, including the NIHSS or MIP images, is better than only PMs.
However, it is not entirely clear if including both NIHSS and MIP images improves the models compared to only including NIHSS in addition to PMs.
Based on the results presented in Table \ref{tab:exps}, we select $\text{SF}_G$(PMs,N) as the proposed model.
One can argue that $\text{SF}_U$(PMs,N) yields similar results, but $\text{SF}_G$(PMs,N) is chosen because of its lower $\Delta V$ for the LVO group, high Dice coefficient over the entire dataset, and satisfactory results for all the other metrics.
Furthermore, \textit{Exp-3} favored the proposed slow fusion approach over the two early fusion approaches, $\text{EF}_G$(PMs,N) and $\text{EFI}_G$(PMs,N).

Results from the 33 randomly selected patients constituting the test set using our selected proposed model (Table \ref{tab:iov}) show an average Dice coefficient of 0.34 for core and 0.63 for penumbra over the entire test set and, on the LVO set, 0.81 and 0.52, respectively.
These results present higher or analogous values compared to the inter-observer variability in most of the metrics, regardless of the stroke's severity.
A separate comparison of the predicted outputs with both the neuroradiologists' annotations demonstrates that the proposed architecture can achieve high statistic values, regardless of the neuroradiologist.
This achievement can be considered valuable throughout the first stages of an AIS.

Predictions from six brain slices by six patients in the test set are shown in Fig. \ref{fig:plots}.
Results display comparable regions as the ground truth images, both with LVO and Non-LVO groups, showing high Dice coefficient and promising results with the model.
However, the third example in Fig. \ref{fig:plots} shows false-positive regions in the brain: this is possibly due to artifacts present in the generated color-coded PMs.
We have noticed similar false-positive with all the other architectures and hyper-parameters as well.
These false-positive might be avoided using 4D raw CTP datasets instead of the pre-generated PMs.

Several researchers have proposed methods with promising results to segment the ischemic core \cite{abulnaga2018ischemic,clerigues2019acute,kasasbeh2019artificial}. They all use PMs derived from CTP studies for their architectures.
Nevertheless, their only focus was to segment the ischemic core without considering the penumbra; thus, excluding a critical aspect for medical treatment decisions.
Furthermore, there was no differentiation in the severity level of AIS for the patients involved in their research.

\section{Conclusion}
The ability to achieve comparable results with two expert neuroradiologists for all the segmented classes (core and penumbra) is valuable to neuroradiologists during the first stages of an ischemic stroke.
The proposed model might be a supportive tool that can help doctors to make treatment decisions.

\bibliographystyle{abbrvnat}
\bibliography{strings}

\end{document}